\documentclass[pra,aps,twocolumn,amsmath]{revtex4-1}

\usepackage{graphicx}
\usepackage{hyperref}
\hypersetup{colorlinks=true, citecolor=blue, urlcolor=blue, linkcolor=blue}

\begin{document}

\title{Nonlocal thermoelectric effects in high-field superconductor-ferromagnet hybrid structures}

\author{J. Heidrich}
\affiliation{Institute of Nanotechnology, Karlsruhe Institute of Technology, Karlsruhe, Germany}
\author{D. Beckmann}
\email[e-mail address: ]{detlef.beckmann@kit.edu}
\affiliation{Institute of Nanotechnology, Karlsruhe Institute of Technology, Karlsruhe, Germany}

\date{\today}

\begin{abstract}
We report on the experimental observation of nonlocal spin-dependent thermoelectric effects in superconductor-ferromagnet multiterminal structures. Our samples consist of a thin superconducting aluminum wire with several ferromagnetic tunnel junctions attached to it. When a thermal excitation is applied to one of the junctions in the presence of a Zeeman splitting of the density of states of the superconductor, a thermoelectric current is observed in remote junctions at distances exceeding $10~\mathrm{\mu m}$. The results can be explained by recent theories of coupled spin and heat transport in high-field superconductors.
\end{abstract}

\maketitle

\section{Introduction}

Superconducting spintronics aims at utilizing the spin degree of freedom of either Cooper pairs or quasiparticles to implement functional electronic devices \cite{eschrig2011,linder2015,beckmann2016,bergeret2018}. Recently, large spin-dependent thermoelectric effects have been predicted \cite{machon2013,ozaeta2014} and observed \cite{kolenda2016,kolenda2016b,kolenda2017} in high-field superconductor-ferromagnet tunnel junctions. These thermoelectric effects are predicted to lead to exceptional thermoelectric figures of merit $ZT\sim 40$ in optimized structures \cite{linder2016}, and can be potentially applied in high-resolution thermometers \cite{giazotto2015b}, radiation detectors \cite{heikkila2018} and coolers \cite{rouco2018}. Spin-dependent thermoelectric effects in these structures are linked to coupled long-range spin and heat transport \cite{silaev2015b,bobkova2015a,bobkova2016,krishtop2015}, which can be driven either by voltage or thermal bias \cite{machon2013}. Voltage-driven long-range spin transport has been observed experimentally \cite{huebler2012b,quay2013,wolf2013}, but a direct experimental proof of the thermal nature of spin tranport in these experiments is still missing. Coupled spin and heat transport is predicted to lead to nonlocal thermoelectric effects in multiterminal superconductor-ferromagnet hybrid structures, where thermal bias on one junction produces a thermoelectric current in a remote junction \cite{machon2013}. Here, we report the first experimental observation of these nonlocal thermoelectric effects.

\section{Experiment}

\begin{figure}[bt]
\includegraphics[width=\columnwidth]{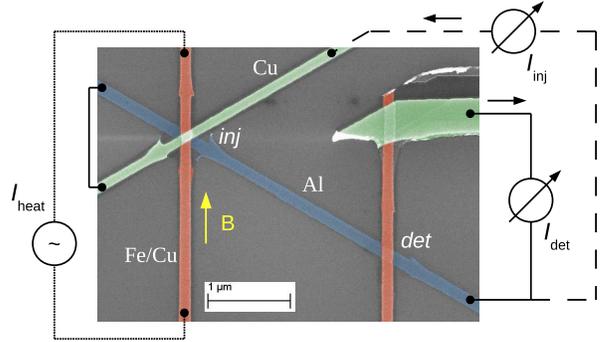}
\caption{\label{fig_sample}
(color online)
False-color scanning electron microscopy image of sample 1 with measurement configurations for the local ($I_\mathrm{inj}$) and nonlocal ($I_\mathrm{det}$) thermoelectric currents.}
\end{figure}

Figure \ref{fig_sample} shows a false-color scanning electron microscopy image of sample 1. The sample has been fabricated by electron beam lithography and shadow evaporation techniques. The sample consists of an aluminum wire of about 15~nm thickness, which was evaporated first, and subsequently oxidized in a partial oxygen atmosphere to create a thin aluminum oxide tunnel barrier. Superimposed are two or more iron wires of about 12~nm thickness, which form spin-polarized tunnel junctions to the aluminum to serve as injector or detector junctions for nonlocal transport experiments. The iron wires are backed by 15-20~nm copper to reduce resistance. An additional copper wire of 50~nm thickness is superimposed onto one of the iron wires (injector) under a different angle to serve as additional measurement probes. Four samples of slightly different design were measured. Sample 1 (shown) had two ferromagnetic junctions (injector and detector), sample 2 had a ferromagnetic and a normal junction, and both could be used either as injector or detector. Samples 3 and 4 had one injector and five detector junctions, at distances $d$ ranging from 1.6 to 12 $\mathrm{\mu m}$ from the injector. An overview of sample parameters is given in Table~\ref{tab:params}.

For transport measurements, the samples were mounted into a shielded box attached to the mixing chamber of a dilution refrigerator, with a magnetic field $B$ applied in the sample plane along the direction of the iron wires. Local and nonlocal differential conductance measurements were performed using standard low-frequency ac lockin techniques. The measurement scheme for the local and nonlocal thermoelectric effects are indicated in Fig.~\ref{fig_sample}. In each case, an ac heater current was applied to the iron wire of the injector junction, creating a thermal excitation across the junction via Ohmic heating. The local thermoelectric current flowing into the aluminum was measured by second harmonic detection using one of the copper leads, as indicated by the dashed line. In the nonlocal configuration, the current flowing out of the aluminum was measured, as indicated by the solid line, and an additional short was placed between the injector and the aluminum wire to ensure $V=0$ across the injector junction (in the local configuration, the low input impedance of the current amplifier ensures $V=0$).

\section{Model}

The spectral properties of the superconductor are calculated using the standard model of high-field superconductors \cite{maki1964b,meservey1975}, including the effect of the Zeeman energy $E_z=\pm\mu_\mathrm{B}B$, the orbital depairing strength $\zeta=\alpha_\mathrm{orb}/\Delta$, the spin-orbit scattering strength $b_\mathrm{so}=\hbar/3\tau_\mathrm{so} \Delta$, and a phenomenogical Dynes broadening $\Gamma$ \cite{dynes1978}, where $\mu_\mathrm{B}$ is the Bohr magneton, $\tau_\mathrm{so}$ is the spin-orbit scattering time, and we have assumed the free-electron $g$ factor of 2. For field-dependent fits, the pair potential $\Delta$ was calculated self-consistently according to Ref.~\onlinecite{alexander1985}, including the effect of Fermi-liquid renormalization of the effective spin splitting with the Fermi-liquid parameter $G^0$. The latter was found to improve the fits in the vicinity of the critical field. For the self-consistent calculations, the orbital depairing was parametrized by
\begin{equation}
 \frac{\alpha_\mathrm{orb}}{\Delta_0} = \frac{1}{2}\left(\frac{B}{B_\mathrm{c,orb}}\right)^2,
\end{equation}
where $\Delta_0=\Delta(T=0,B=0)$. From this model, we obtain the spin-resolved density of states $N_\pm$ of the superconductor, and the renormalized diffusion coefficent $D_L$ used for the nonequilibrium model (see below). The fits are in general not very sensitive to $b_\mathrm{so}$, $G^0$ and $\Gamma$, and we have chosen $b_\mathrm{so}=0.015$, $G^0=0.25$ and $\Gamma=0.005\Delta_0$ for all fits in the paper. 

To describe local and nonlocal currents under nonequilibrium conditions on an equal footing, we use the quasiclassical distribution functions $f_L$ and $f_T$ throughout the model \cite{schmid1975,silaev2015b,bobkova2016}. In thermal equilibrium, for a conductor held at temperature $T$ and electrochemical potential $\mu$, these are given by $f_L=n_+$ and $f_T=n_-$, where
\begin{equation}
 n_\pm(E,\mu,T) = \frac{1}{2}\left(n_0(E+\mu,T)\pm n_0(E-\mu,T)\right),
\end{equation}
$n_0(E,T)=\tanh(E/2k_\mathrm{B}T)$, $E$ is the energy, and $k_\mathrm{B}$ is the Boltzmann constant. In the following, we only consider the nonequilibrium parts, i.e., $f_L$ implicitly means $f_L-n_0(T)$, where $T$ is the electronic base temperature in the absence of thermal excitation. Throughout the paper, we distinguish the base temperature $T_0$ of the cryostat and the electronic base temperature $T$. $T$ may be increased above $T_0$ due to incomplete filtering of the measurement lines, and is determined by fitting the data. 

We now consider a tunnel junction between a ferromagnet and a superconductor, with normal-state conductance $G$ and spin polarization $P$. The ferromagnet is held at temperature $T+\delta T$ and chemical potential $\mu=eV$ with respect to the superconductor, where $e=-|e|$ is the charge of the electron. The current into the superconductor is then given by \cite{ozaeta2014,silaev2015b,bobkova2016}
\begin{multline}
 I(V,\delta T) = \frac{G}{e}\int_0^\infty \left[N_0(E)\delta f_T(V,\delta T)\right.\\
 -\left.PN_z(E)\delta f_L(V,\delta T)\right]dE,
  \label{eqn_local}
\end{multline}
where $\delta f_T$ and $\delta f_L$ are the differences of the distribution functions across the junction, and $N_0=(N_+ + N_-)/2$ and $N_z=(N_+ - N_-)/2$ are the average and difference of the spin-resolved densities of states, respectively. Note that the thermoelectric contribution to the tunnel current (for $V=0$ and $\delta T\neq 0$) is given by the second term of the integrand of Eq.~(\ref{eqn_local}).

\begin{figure}[bt]
\includegraphics[width=\columnwidth]{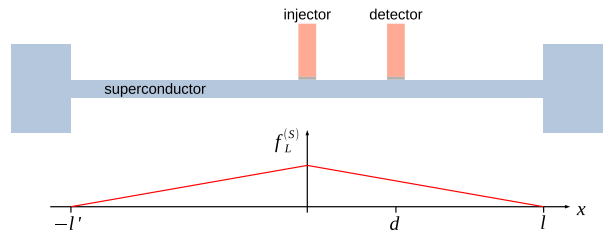}
\caption{\label{fig_model}
(color online) Sketch of the nonequilibrium model. A superconducting wire along the $x$ axis is connected to two reservoirs at each end (at $x=-l'$ and $x=l$). An injector and detector tunnel junction are placed at $x=0$ and $x=d$, respectively. Nonequilibrium is modeled by the energy-mode distribution function $f^{(S)}_L(x)$.}
\end{figure}

To describe the nonlocal conductance, we restrict ourselves to the most simple model that captures the basic physics. The system considered is shown schematically in Fig.~\ref{fig_model}. A superconducting wire along the $x$ axis is attached to two equilibrium reservoirs at $x=-l'$ and $x=l$. An injector tunnel junction is placed at $x=0$, and a detector is placed at $x=d$. The total length of the wire is $l+l'$. For our samples, $l\approx l'\approx 20~\mathrm{\mu m}$. Of the four possible nonequilibrium modes, we consider only $f_L$, and neglect inelastic scattering \cite{silaev2015b}. In this limit, nonequilibrium injection is balanced by the diffusion of the quasiparticles into the reservoirs, and the solution is a linear function of $x$. It is given by
\begin{equation}
 f^\mathrm{(S)}_L(x) = G_\mathrm{inj}R \frac{N_0 f^\mathrm{(inj)}_L - P_\mathrm{inj} N_z f^\mathrm{(inj)}_T}{D_L+G_\mathrm{inj}R N_0} \left(1-\frac{x}{l}\right)\label{eqn:fL}
\end{equation}
for $0<x<l$. Here $R$ is the normal-state resistance of the two branches of the superconducting wire to the left and to the right of the injector in parallel. $D_L$ is the spectral diffusion coefficent for the longitudinal mode, which is extracted from the same model as the densities of states. $f^\mathrm{(inj)}_{L,T}$ are the distribution functions in the injector junction. The current flowing out of the detector junction is then given by
\begin{equation}
 I_\mathrm{det} = -G_\mathrm{det}P_\mathrm{det}\mu_z,\label{eqn:Idet}
\end{equation}
where
\begin{equation}
 \mu_z = \frac{1}{e}\int_0^\infty N_z f^\mathrm{(S)}_L(d) dE\label{eqn:muz}
\end{equation}
and we assume that the detector distribution is at equilibrium ($f_{L,T}^\mathrm{(det)}=0$).

The nonequilibrium distribution $f^\mathrm{(S)}_L(x=0)$ is not necessarily small compared to the injector distribution $f^\mathrm{(inj)}_L$, in particular for the thermoelectric measurements. Therefore, in all fits of the conductance and thermoelectric effect shown in this paper, we set $\delta f^\mathrm{(inj)}_L=f^\mathrm{(inj)}_L-f^\mathrm{(S)}_L(x=0)$.

So far, the model completely neglects inelastic scattering. While a full treatment of electron-electron and electron-phonon scattering is beyond the scope of this paper, we can still include thermalization of quasiparticles by electron-electron scattering in a phenomenogical way: Following Ref.~\onlinecite{bobkova2016}, we define an effective nonequilibrium temperature $T_\mathrm{S}$ of the superconductor by setting the excess energy equal to the one given by $f^\mathrm{(S)}_L$, {\em i.e.}, by setting
\begin{equation}
 \int_0^\infty N_0 E \left(f^\mathrm{(S)}_L+n_0(T)-n_0(T_\mathrm{S})\right) dE=0 \label{eqn:defts}
\end{equation}
and solving for $T_\mathrm{S}$.

\section{Results}

\begin{figure}[bt]
\includegraphics[width=\columnwidth]{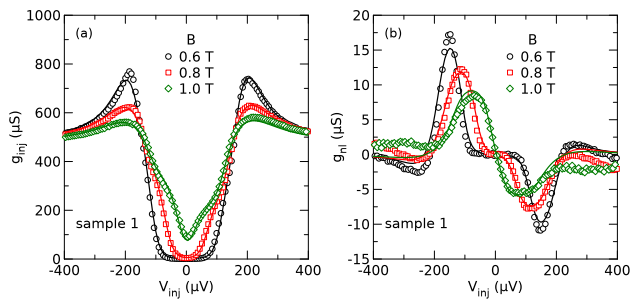}
\caption{\label{fig_conductance}
(color online) (a) Local differential conductance $g_\mathrm{inj}$ and (b) nonlocal differential conductance $g_\mathrm{nl}$ as a function of bias voltage $V_\mathrm{inj}$ for different magnetic fields at base temperature $T_0=50~\mathrm{mK}$.}
\end{figure}

To characterize our samples, we measured the local differential conductance $g=dI/dV$ for each junction. As an example, we show the conductance $g_\mathrm{inj}$ of the injector junction of sample 1 measured at different magnetic fields $B$ in Fig.~\ref{fig_conductance}(a). The conductance has the typical shape of the BCS density of states, and at high fields the Zeeman splitting and the asymmetry due to the spin polarization of the junction are visible. Lines are fits with our model, from which we extract the spin polarization.

Figure~\ref{fig_conductance}(b) shows the nonlocal differential conductance $g_\mathrm{nl}=dI_\mathrm{det}/dV_\mathrm{inj}$ measured simultaneously with the local conductance in Fig.~\ref{fig_conductance}(a). The data exhibit two broad peaks of opposite sign in the bias range of the Zeeman splitting, as observed earlier \cite{huebler2012b,quay2013}. Model predictions are shown as lines. All parameters for these predictions were determined independently, with no free fit parameters left.

\begin{table}[tb]
\begin{tabular}{cccccccccc}
       & $T_\mathrm{c}$ & $B_\mathrm{c}$ & $G$                & $P$         & $B_\mathrm{c,orb}$ & $G_\mathrm{inj}R$  \\
sample & $(\mathrm{K})$ & $(\mathrm{T})$ & $(\mathrm{\mu S})$ &             & $(\mathrm{T})$     &       \\ \hline
1      &  $1.44$        & $1.32$         & $360-420$          & $0.24-0.28$ & $1.44$             & 0.104 \\
2      &  $1.46$        & $1.43$         & $170-300$          & $0.19$      & $1.51$             & 0.062 \\
3      &  $1.47$        & $1.30$         & $150-190$          & $0.25-0.29$ & $1.43$             & 0.052 \\
4      &  $1.48$        & $1.44$         & $140-150$          & $0.16-0.18$ & $1.52$             & 0.052  \\
\end{tabular}
\caption{Overview of sample and fit parameters. Critical temperature $T_\mathrm{c}$ and critical field $B_\mathrm{c}$ determined from the onset of superconductivity in the conductance measurements. Junction conductance $G$ and spin polarization $P$ extracted from the fits of the conductance spectra. Parameter $B_\mathrm{c,orb}$ extracted from the fits of $I_\mathrm{inj}(B)$, and $G_\mathrm{inj}R$ calculated from $G_\mathrm{inj}$ and the normal-state wire resistance.}\label{tab:params}
\end{table}

\begin{figure}[bt]
\includegraphics[width=\columnwidth]{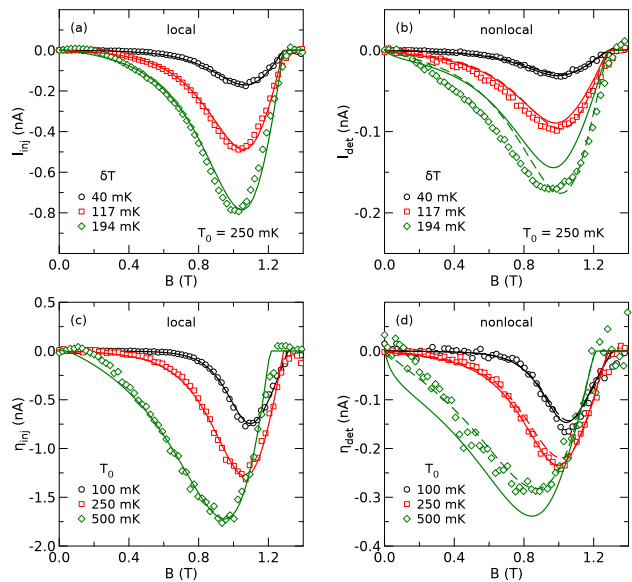}
\caption{\label{fig_te}
(color online) Thermoelectric signal in the local configuration (left) and nonlocal configuration (right) measured on sample 1 under the same experimental conditions.
(a) Local and (b) nonlocal thermoelectric current as a function of applied magnetic field $B$ for different thermal excitations $\delta T$.
(c) Local and (d) nonlocal thermoelectric coefficient $\eta$ as a function of applied magnetic field $B$ for different base temperatures $T_0$.}
\end{figure}

Figure \ref{fig_te} shows an overview of the thermoelectric signals obtained for sample 1. Figure \ref{fig_te}(a) shows the local thermoelectric current $I_\mathrm{inj}$ as a function of magnetic field $B$ for different thermal excitations $\delta T$, measured at a base temperature $T_0=250~\mathrm{mK}$. The signal is zero at zero applied field, and then a negative thermoelectric current develops upon increasing the field. The maximum signal is observed at about $1.1~\mathrm{T}$, and then quickly decreases towards the critical field at about $1.3~\mathrm{T}$. Similar signals have been observed in our previous work \cite{kolenda2016}. The lines in the plot are fits with our model. For these fits, we kept all parameters fixed to the ones determined independently, and used $B_\mathrm{c,orb}$, $T$ and $\delta T$ as free parameters for fitting the data at small excitation. For the larger excitations, only $\delta T$ was allowed to vary.

The nonlocal thermoelectric current measured under the same conditions is shown in Fig.~\ref{fig_te}(b). It exhibits the same qualitative behavior as the local current, but is smaller by about a factor of four. Solid lines are model predictions based on Eq.~(\ref{eqn:fL}), again without free parameters. As can be seen, the agreement is excellent at small excitation, but at larger excitation, the model underestimates the signal. Dashed lines are predictions including thermalization according to Eq.~(\ref{eqn:defts}). They do not differ much at small excitation, but give a slightly better description of the signal for larger excitation (and therefore larger quasiparticle excess energy).

Figures \ref{fig_te}(c) and \ref{fig_te}(d) show the temperature dependence of the local and nonlocal thermoelectric effect, respectively. Thermal excitations were about $50~\mathrm{mK}$. To compare the data for different temperatures, we plot the normalized coefficent $\eta=IT/\delta T$. For fitting the local data, we kept $B_\mathrm{c,orb}$ fixed to the value from the fit at $T_0=250~\mathrm{mK}$, and allowed only $T$ and $\delta T$ to vary. Solid and dashed lines in Fig.~\ref{fig_te}(d) are model predictions without and with thermalization, again without free parameters. There is little difference at low temperatures, but at $T_0=500~\mathrm{mK}$ the thermalized model gives a better fit.

\begin{figure}[bt]
\includegraphics[width=\columnwidth]{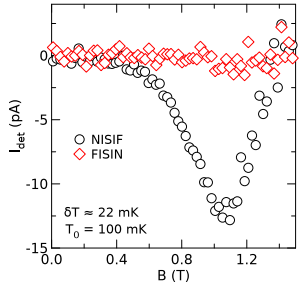}
\caption{\label{fig_xtra}
(color online) Nonlocal thermoelectric current $I_\mathrm{det}$ as a function of applied magnetic field $B$ for two measurement configurations for sample 2. NISIF: normal injector, ferromagnetic detector. FISIN: ferromagnetic injector, normal detector.}
\end{figure}

In Fig.~\ref{fig_xtra}, we compare two different measurement configurations for sample 2. This sample had a ferromagnetic (F) and normal-metal (N) junction, which could be both used as injector or detector. We compare here the configurations with normal injector and ferromagnetic detector (NISIF), and the reverse configuration (FISIN). According to Eq.~(\ref{eqn:fL}), for pure thermal bias, {\em i.e.}, $f^\mathrm{(inj)}_T=0$, the nonequilibrium distribution should not depend on the injector polarization, whereas according to Eq.~(\ref{eqn:Idet}), the detector current should disappear for a normal detector ($P_\mathrm{det}=0$). In agreement with this prediction, we observe a thermoelectric current for the NISIF configuration, but no signal for the FISIN configuration. The same behavior has been observed previously for bias-driven spin injection \cite{wolf2013}.

\begin{figure}[bt]
\includegraphics[width=\columnwidth]{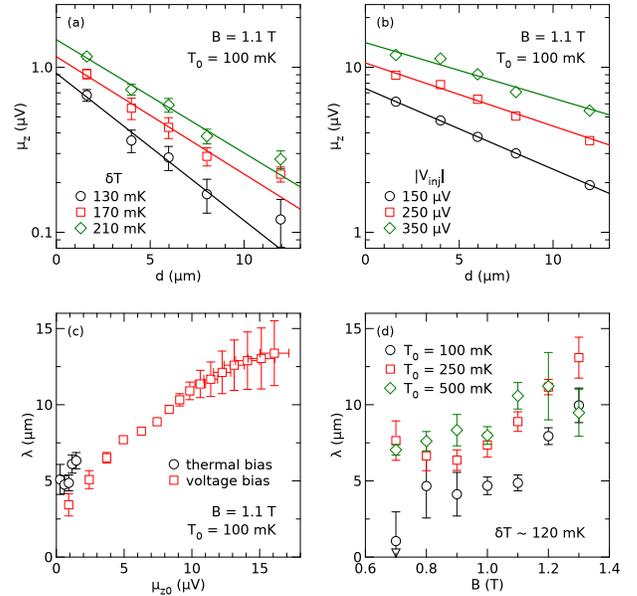}
\caption{\label{fig_dist}
(color online) $\mu_z$ plotted on a logarithmic scale as a function of contact distance $d$ for (a) thermal bias and (b) voltage bias. Lines are exponential fits according to Eq. (\ref{eqn:expfit}). (c) relaxation length $\lambda$ extracted from these fits as a function of signal amplitude $\mu_{z0}$. (d) relaxation length $\lambda$ as a function of magnetic field $B$ for different base temperatures $T_0$. All data are from sample 4.}
\end{figure}

In Fig.~\ref{fig_dist}, we show the dependence of the nonlocal signal on contact distance $d$. All data are taken from sample 4, where we collected the most extensive data set. Similar results were found for sample 3. To eliminate small junction-to-junction variations of the detectors, we plot $\mu_z=-I_\mathrm{det}/G_\mathrm{det}P_\mathrm{det}$. 
Figure~\ref{fig_dist}(a) shows $\mu_z$ vs. $d$ extracted from the nonlocal thermoelectric effect for different thermal excitation $\delta T$. Data are averaged  over a field interval of $\pm 50~\mathrm{mT}$ around $B=1.1~\mathrm{T}$, where the signal maximum occured for this sample. Since our simple model Eq.~(\ref{eqn:fL}) neglects all relaxation processes, it does not capture the decay of the nonlocal signals as a function of $d$ realistically. We therefore fit the data phenomenogically with an exponential decay
\begin{equation}
 \mu_z(d) = \mu_{z0}\exp(-d/\lambda).\label{eqn:expfit}
\end{equation}
Figure~\ref{fig_dist}(b) shows $\mu_z$ as a function of $d$ for bias-driven spin injection at the same temperature and field for different bias voltages $|V_\mathrm{inj}|$ (data are averaged for positive and negative bias). The signal is larger by about a factor of ten, reflecting the much stronger nonequilibrium conditions imposed by voltage bias (using $eV_\mathrm{inj}=k_\mathrm{B}\delta T$, $V_\mathrm{inj}=20~\mathrm{\mu V}$  corresponds to about  $\delta T=200~\mathrm{mK}$). To directly compare voltage and thermal bias, we plot the decay length $\lambda$ obtained from the fits as a function of $\mu_{z0}$ in Fig.~\ref{fig_dist}(c). In either case, the relaxation length is about $5~\mathrm{\mu m}$ for weak bias, and increases with increasing bias. The same qualitative behavior has been found in our previous work on bias-driven spin injection \cite{wolf2014c}. In Fig.~\ref{fig_dist}(d), we finally show an overview of the relaxation length of the nonlocal thermoelectric signal for different applied fields $B$ and base temperatures $T_0$. The relaxation length increases with increasing field, as observed previously \cite{huebler2012b,wolf2013}. There is also an increase with temperature, which was not observed in the bias-driven case \cite{wolf2013}.

\section{Conclusion}

We have reported the first experimental observation of nonlocal spin-dependent thermoelectric effects in superconductor-ferromagnet hybrid structures. The results can be explained by theoretical models based on coupled spin and heat transport, and the decay length is consistent with the relation length for bias-driven spin transport. More extensive theoretical modeling may provide insights into inelastic relaxation mechanisms in high-field superconductors, and in particular distinguish electron-electron and electron-phonon scattering. Future investigations could probe nonlocal Peltier effects, and generalized nonlocal Onsager relations.

\acknowledgements{This work was supported by the DFG under Grant No. BE-4422/2-1.}

\bibliography{nonlocalte}

\end{document}